\documentstyle[preprint,eqsecnum,aps]{revtex}

\tightenlines

\def\gtorder{\mathrel{\raise.3ex\hbox{$>$}\mkern-14mu
             \lower0.6ex\hbox{$\sim$}}}
\def\ltorder{\mathrel{\raise.3ex\hbox{$<$}\mkern-14mu
             \lower0.6ex\hbox{$\sim$}}}

\begin{document}
\draft
\title{Experimental hints of Gravity in Large Extra Dimensions?
\footnote{This essay received an "honorable mention" in the 2001 Essay
Competition of the Gravity Research Foundation.}
}
\author{Steinn Sigurdsson}
\address{525 Davey Laboratory,
Department of Astronomy \& Astrophysics,\\ 
Pennsylvania State University, University Park, Pa 16802\\
steinn@astro.psu.edu
}
\maketitle
\begin{abstract}
Recent conjectures suggest
the universe may have large extra dimensions, through which gravity 
propagates. This implies gross departures from Newton's law of gravity
at small length scales. 
Here I consider some implications for particle dynamics on scales
comparable to the compactification radius, $R_c \ltorder 1$ mm.
During planet formation, coalescence of micron sized dust grains to
planetesimals is a rate critical step. 
Blum et al (2000) found dust grain aggregates form low fractal
dimension structures in microgravity, consistent with high angular
momentum coalescence.
I consider the effects of non-Newtonian gravity on dust aggregation on
scales less than $R_c$ and show they naturally coalesce into low dimensional
structures with high specific angular momentum.
We infer $R_c \approx 80$ microns.
\end{abstract}


\vfill\eject

\narrowtext
\section{
}
\label{sec:level1}

Recent conjectures have postulated that the hierarchy problem
in physics may be resolved if two (or more) of the extra dimensions
postulated by extensions of the standard model of particle theory,
are compactified on mesoscopic scales - with effective radii
much larger than the Planck scale \cite{ark98,ras99}.
A particularly interesting possibility is in the case of $n=2$
mesoscopic compactification, in which case the implied scale, $R_c$,
for the extra dimensions is of the order of 0.1 mm.
In the simplest theory, the standard model gauge fields are
restricted to (or near) the 3--dimensional brane on which we
normally live, and only gravity propagates into the bulk of the large extra dimensions (LEDs).
The resulting theory has a number of interesting astrophysical implications \cite{ark99}.  

An immediate implication of LEDs is that Newton's law fails on small scales, and
is replaced by an effective potential gradient

\begin{equation}
\nabla \Phi(r) = -(n+1){ {m_1m_2}\over {M_{pl-n}^{n+2}} } { {1}\over {r^{n+2}} } \quad r \ll R_c
\end{equation}
where $n$ is the number of large extra dimensions, and $M_{pl-n}$ is the higher
dimensional Planck mass, implying an effective 
4-D Planck mass $M_{pl} \sim M_{pl-n}^{1+n/2} R_c^{n/2}$ \cite{ark98}.
The laboratory experimental
constraints on deviations from Newton's law on scales less than 1 cm \cite{lcp99,hoy01} are
weak, so the
conjecture is not directly excluded by direct experiments, although recent experiments
have constrained $R_c < 218 \mu m$ \cite{hoy01}, and future experiments should detect $R_c$.
If correct, LEDs have many implications for physics
on different scales, some of which will be tested in the near future. 

It is generally accepted that planets form through an aggregation of smaller
grains. The protoplanetary disk is expected to contain sub--micron sized dust
(and ice) grains which segregate through sedimentation to the mid--plane of the disk.
The disk may be turbulent on some scales, and in general there is
drag on the grains due to the gas orbiting at slightly sub--Keplerian velocities.
The basic picture for planet formation has planetesimals form by dust aggregation,
with runaway accretion onto the largest planetesimals once they reach mesoscopic sizes
and the gravitational field of the largest planetesimals dominates the local potential
\cite{lis87,rup91,wei97,rud99}. The accretion of planetesimals builds masses up to
several times the mass of the Earth, at which point direct accretion of gas is thought
to runaway to produce the Jovian planets.
A problem is presented, because observational evidence and indirect theoretical
arguments require that the time scale for dust grains to assemble into large planetesimals must
be very short, of order $10^6$ years or shorter. 

A critical step in this process is the growth from sub--micron sized dust grains
to $\sim 1$ cm (see eg review by Ruden \cite{rud99}), 
at which point radial drag allows grains to effectively sweep
up a large volume rapidly, and to grow to the point where the self--gravity of
the largest planetesimals produces rapid coalescence. 
In order for small grains to grow rapidly enough, it is necessary
to postulate that they form loose aggregates with fractal dimension $D_f\leq 2$ in
order for the grains to effectively sweep up smaller grains \cite{wei97,wb98}. 
Yet if the grain--grain
velocity is large enough for significant growth to occur, we expect compactification
or fragmentation during grain collisions, which slows down growth, as the geometric
cross--section of the grains is reduced.

Recently Blum et al (2000) experimentally measured aggregation of micron sized dust
grains in microgravity, in the CODAG experiment on STS-95. Rather surprisingly, they
found dust grain growth up to $\sim 50 \mu m$ producing branched linear structures
with $D_f \sim 1.3$, much lower than predicted. The implied grain-grain coalescence
time scale is then almost independent of grain mass ($\tau \sim m^{0.06}$), the dust mass
function is dominated by large grains, and aggregation processes are dominated
by large grains.

To explain this process, Blum et al impose an ad hoc cut-off on aggregration, where
sticking is restricted to high impact parameter collisions ($b/a > 0.65$, for impact
parameter $b$ on grain radius $a$). They conjecture that thermal rotation biases 
grain impact to large impact parameters. An alternative conjecture is that non-Newtonian
gravitational attraction biases collisions to large effective impact parameters for $a < R_c$.

The existence of large extra dimensions, in which gravity propagates, changes
the dynamics of grain--grain interactions at scales smaller than $R_c$ \cite{sig00}.
In the Newtonian regime, for grains size $a$ and density $\rho \sim 1$, the
surface escape velocity $v_N \approx 2a \sqrt{G\rho }$ and the self--gravity
of a grain is irrelevant for plausible grain dispersions for sizes less than $\sim 100$ m.
Grain coagulation requires high grain velocity dispersion so small grains can have large
collision rates, implying large velocity gradients or small scale turbulence.
Countering this is the problem that at relative velocities $\gtorder 100\ {\rm cm\, s^{-1}}$,
experiments show that grains do not stick and growth is inhibited \cite{wb98}.

A lower bound on local dispersion is set by the Brownian motion
induced dispersion, $v_B \sim 10^{-5} \ {\rm cm\, s^{-1}} \sqrt{(T/100)}/\sqrt{m_{-3}}$,  
where $m_{-3}$ is the grain mass in milligrams. A $100 \mu m$ grain has a mass of about $0.1 m_{-3}$
if it is compact. 
The smallest grains have high dispersion
due to Brownian motion, but intermediate sized grains, of order 100 microns, 
have low dispersion and low number densities. 

The gas-grain response time is a few msec \cite{blu00}, so a dust grain approaching within
$R_c \sim 100 \mu m$ of an aggregate, will explore a range of relative velocities
while it traverses the region of non-Newtonian attraction. This favours grain growth if there
is no small scale turbulence and gas velocity gradients are small.

With LEDs, the escape velocity of grains smaller than $R_c$ is independent of the
grain size and is just $v_6 \approx v_N(R_c) \sim 10^{-4}\ {\rm cm\, s^{-1}}$. 
Further, orbits in $1/r^3$ potentials are unstable, so any grains approaching
within $R_c$ of each other, with instantaneous relative velocities less than $v_6$ become bound 
and must coalesce. A non-circular orbit in a $r^{-3}$ potential spirals into contact
on a dynamical time, so any particle random walking in velocity so that it becomes
bound to the aggregate cluster, immediately coalesces with the cluster.
For sub-millimeter sized grains, this increases the collision cross--section 
by $\sim 2$ orders of magnitude,
at low velocity dispersion,
and makes the issue of sticking and compactification in collisions moot, {\it if} 
the grain--grain velocities are low.
Thus in LED modified gravity, grains in lower density dust can
coagulate rapidly, if the grain--grain velocity dispersion is {\bf small} on scales $\ltorder R_c$.
In disks, gas drag imposes a size dependent velocity gradient on dust grains, but 
the grain--grain dispersion is in general much smaller than the differential
velocity between grains and gas, in the CODAG experiment 
the local grain--grain velocity dispersion
is small and dominated by Brownian motion.

More importantly, the coalescing grains have much larger relative angular momentum than
in the Newtonian case. A typical grain will have an orbital eccentricity of order $0.7$ 
when it becomes bound, and enters
the non-Newtonian regime at apocenter $r_a \sim R_c$. So the specific angular momentum of the
grain is $l \sim {1\over 2} v_{c-Newtonian}(R_c)\times R_c \sim G\rho a^3 \sqrt{R_c}$.
If the distribution of impact parameters is uniform in area, then the excess angular momentum
observed by Blum et al, is consistent with $R_c \approx 80 \mu m$.   

Establishing the conditions necessary for rapid planet formation,
as required by observations, has required some considerable fine tuning of
the disk initial conditions in the models.
With the modifications to Newtonian gravity that follow from theories of large extra
dimensions, the problem of rapid coalescence at the smallest scales becomes simpler.
The stronger gravitational force and resulting orbit instability on scales smaller than
the effective compactification radius, allow non--Newtonian gravity to dominate 
the particle dynamics of this critical early stage
of planet formation, allowing rapid grain growth
precisely in those cases in which time scales for Newtonian particles
to coalesce become prohibitively long.
In principle the efficient coalescence of micron sized grains 
through non--Newtonian gravitational forces can be tested directly
in microgravity experiments. The recent experiment of Blum et al is
consistent with $R_c \sim 80$ microns, but it is of course quite possible other, Newtonian,
effects are the cause of the anomalous dust grain aggregation observed.

IF the orbit instability inside $R_c$ is the cause of the anomalous grain growth, then
first order estimates suggest $R_c \approx 80$ microns, the corresponding 
unification scale for $n=2$ extra dimensions is about $10 TeV$. This should be testable,
both through direct measurements of deviation from Newtonian gravity on small length scales;
through saturation of grain aggregation at $a \gtorder R_c$, and a transition to grain fractal
dimension $D_f (a>R_c) \sim 2$; and, of course, through direct observation of anomalous particle collisions
cross-sections at $\sim 10$ TeV.

\acknowledgments I would like to thank Sterl Phinney, Neil Cornish and Craig Hogan for helpful
conversations, and to the Aspen Center for Physics for hospitality.


\begin{references}

\bibitem{ark98}Arkani--Hamed, N., Dimopoulos, S. \& Dvali, G., Phys. Lett. B, 429, 263--272, 1998

\bibitem{ark99}Arkani--Hamed, N., Dimopoulos, S. \& Dvali, G., Phys. Rev. D, 59, 086004, 1999

\bibitem{blu00}Blum, J. et al, Phys. Rev. Lett., 85, 2426, 2000

\bibitem{hoy01}Hoyle, C.D., Schmidt, U., Heckel, B.R., Adelberger, E.G., Gundlach, J.H., Kapner, D.J., \& Swanson, H.E., Phys. Rev. Lett., 86, 1418, 2001

\bibitem{lis87}Lissauer, J.J., Icarus, 69, 249--265, 1987

\bibitem{lcp99}Long, J.C., Chan, H.W. \& Price, J.C., Nuc. Phys. B, 539, 23--34, 1999

\bibitem{ras99}Randall, L. \& Sundrum, R., Phys. Rev. Lett. 83, 4690--4693, 1999

\bibitem{rup91}Ruden, S.P. \& Pollack, J.B., Astro. Phys. J., 375, 740--760, 1991

\bibitem{rud99}Ruden, S.P., in {\it The Origin of Stars and Planetary Systems}, Conf. Proc., p. 643, Kluwer Academic, eds. Lada \& Kylafis, 1999

\bibitem{sig00} Sigurdsson, S., ``Essays on Gravitation'' competition, 2000

\bibitem{wei97}Weidenschilling, S.J., in {\it From Stardust to Planetesimals}, Conf. Proc. PASP, v. 122, p. 281--311, 1997, eds. Pendleton \& Tielens

\bibitem{wb98}Wurm, G. \& Blum, J., Icarus, 132, 125--136

\end{references}
\end{document}